\documentclass[fleqn,twoside]{article}
\usepackage{espcrc2,epsfig}

\usepackage{graphicx}
\usepackage[figuresright]{rotating}

\newcommand{\AmS}{{\protect\the\textfont2
  A\kern-.1667em\lower.5ex\hbox{M}\kern-.125emS}}

\title {QCD Critical Region and Quark Gluon Plasma
from an Imaginary $\mu_B$} 

\author{Massimo D'Elia\address {Dipartimento di Fisica dell'Universit\`a 
di Genova and INFN, I-16146, Genova, Italy }%
        \thanks{Electronic address: {\tt delia@ge.infn.it}} and
Maria-Paola Lombardo\address {INFN-Laboratori Nazionali di Frascati, 
I-00044, Frascati(RM), Italy}
        \thanks{Electronic address: {\tt lombardo@lnf.infn.it}} }
       
\begin{document}

\begin{abstract}
We discuss the imaginary chemical potential approach to the
study of QCD at nonzero temperature and density, 
 present results for the four flavor model in the different 
phases and show that this method is ideally suited
for a comparison between lattice data and phenomenological models.

\vspace{1pc}
\end{abstract}
\maketitle

\section{QCD and a complex $\mu_B$}
Results from simulations with an imaginary chemical potential
can be analytically continued to a real chemical potential,
thus circumventing the sign problem \cite{Hart:2000ef},
\cite{deForcrand:2002ci} \cite{D'Elia:2002gd}. 
In practice, the analytical
continuation is carried out along one line in the complex
$\mu$ plane: first along the
imaginary axes, and then along the real one. It is then
meaningful to map  this path in the complex $\mu^2$ plane: because
of the symmetry property $Z(\mu)=Z(-\mu)$ this can be achieved 
without losing generality. In the complex $\mu^2$ plane
the partition function is real for real values of the external parameter
$\mu^2$, complex otherwise: the situation resembles
that of ordinary statistical models in an external field. Hence,
the analyticity of the physical observables \cite{Lombardo:1999cz} 
as well as that of the critical line 
\cite{deForcrand:2002ci} follows naturally.

The phase diagram in the temperature, (real) $\mu^2$ plane is
sketched in Fig.1, where we omit the superconducting and the color flavor
locked phase, which (unfortunately) play no r\^ole in our discussion.
The region accessible to numerical simulations is the one with
$\mu^2 \le 0$: at a variance with other approaches to finite density
QCD   which only use information at $\mu=0$ \cite{Fodor:2001au}
\cite{Allton:2003vx} \cite{Crompton:2001ws}
the imaginary chemical potential method exploits the entire halfspace.
And we will also argue that there are physical questions which can 
be addressed without analytic continuation.

\begin{figure}[htb]
\psfig{file=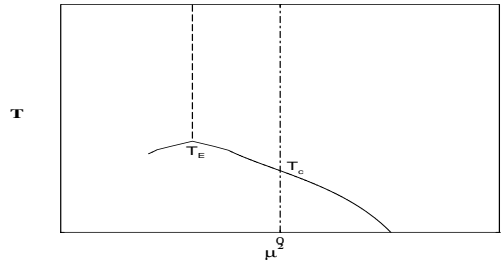,width=6.5 truecm,height=3.5 truecm}
\caption{Sketch of the phase diagram in the $\mu^2,T$ plane:
the solid line is the chiral transition, the dashed
line is the Roberge Weiss transition. Simulations can be carried out at
$\mu^2 \le 0$ and results continued to the
physical domain $\mu^2 \ge 0$}
\end{figure}
\begin{figure}[htb]
\psfig{file=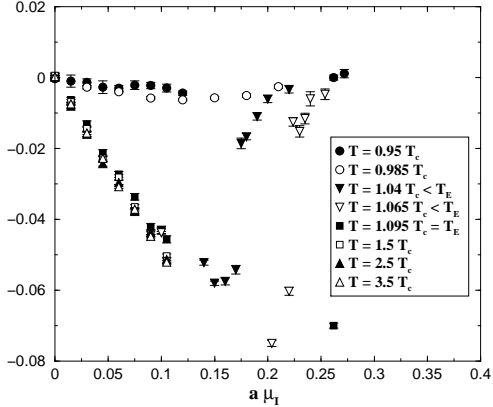,width=6.5 truecm}
\caption{Number density as a function of $\mu_I$: note the smooth behavior
in the hadronic phase, consistent with the hadron resonance gas model,
the chiral transition in the Roberge Weiss region, the rapid increase
in the plasma phase, approaching a nearly free quark behavior.}
\end{figure}
\begin{figure}[htb]
\psfig{file=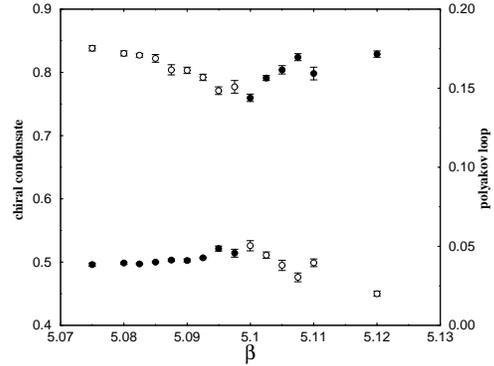,width=6.5 truecm}
\caption{Correlation between $<\bar\psi\psi>$ and Polyakov loop
at $\mu_I = 0.15$, demonstrating the correlation of chiral
and deconfining transition at nonzero baryon density.}
\label{fig:largenenough}
\end{figure}
\begin{figure}[htb]
\psfig{file=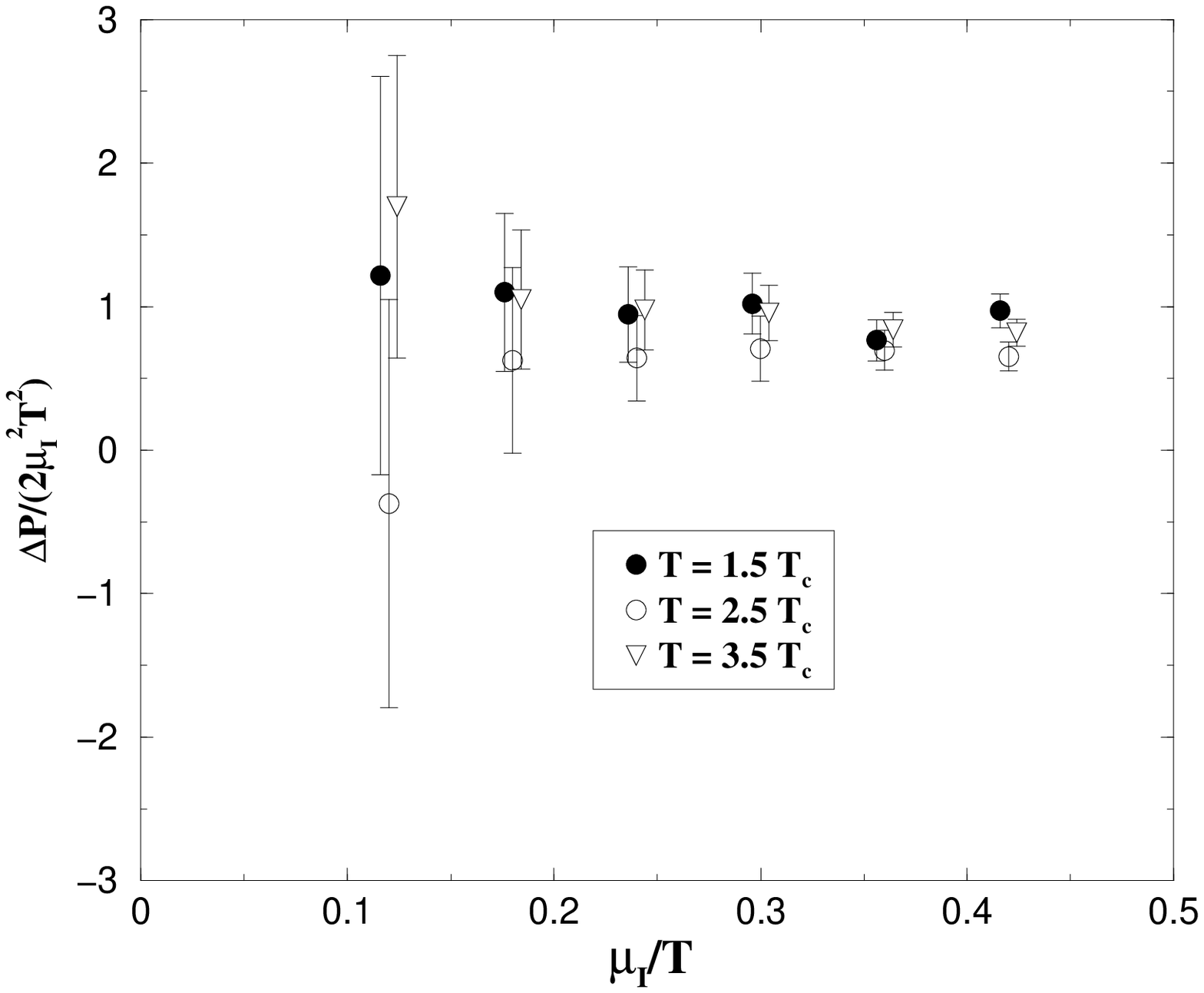,width=6.5 truecm}
\caption{The effective prefactor plot (see text).}
\end{figure}

\section{The Critical Line}

Also in  this case the consideration 
of the $T,\mu^2$ plane helps the analysis.
Model analysis suggests the following parametrization, confirmed by 
numerical results:
\begin{equation}
(T+aT_c)(T - T_c) + k \mu^2  = 0 ,\;\;\;\; k > 0
\end{equation}
It encodes reality for real $\mu^2$,
contains the physical scale $T_c$ , is dimensionally consistent, gives
$T(\mu = 0) = T_c$, $T(\mu \neq 0) < T_c$ .
We refer to Section IV of Ref.~\cite{D'Elia:2002gd}
for our results on the critical line in the four flavor model,
and their discussion in terms of model calculations.
Here it suffices to remind ourselves that the second order approximation
turns out to be adequate, and that the fourth order corrections
were found to be consistent with zero within errors.

\section{Hadronic Phase: $T < T_c$}

In this region observables are a continuous and periodic function
of $\mu_I/T$ , analytic continuation
in the $\mu^2 > 0$ half plane is always possible, but
interesting only when $\chi_q(\mu=0,T) > 0$.

The analytic continuation of an observable $O$ is valid till $\mu < \mu_c(T)$,
where $\mu_c(T)$ has to be measured independently.
The value of the analytic continuation at
$\mu_c$, $O(\mu_c)$ defines the discontinuity
at the critical point, or, equivalently, the critical
value of, say, the number density. In turns, this
allows the identification of the order of the phase transition.

For observables which are even/odd in the chemical potential, 
$O_{e/o}$,  we have considered a  Fourier series,
observing that the first cosine/sine terms
suffice to parametrize the data \cite{D'Elia:2002gd}. 
This has been confirmed in  \cite{Karsch:2003zq},
where this result has been intepreted within the framework of the
hadron resonance gas model.

\section{Roberge Weiss Regime: $T_C < T < T_E$}

The analytic continuation is valid till $\mu = \infty$ 
but the interval accessible to the simulations at
imaginary $\mu$ is small, as
simulations in this area hits the chiral critical
line for $\mu^2 < 0$.

The bright side of this is that the  nature of the critical line
can then be studied without need for analytic continuation.
In Fig. 3 we show the clear correlation between the Polyakov Loop
and the chiral condensate at $\mu=0.15$. The correlation between
chiral and deconfining transition persists at nonzero chemical
potential. 

It is also of interest to note that the non--applicability of perturbation
theory in this region is almost a theorem: indeed the analytic continuation
of the polynomial predicted 
by perturbation theory for positive $\mu^2$ 
would never reproduce the correct critical behavior
at the second order phase transition for $\mu^2 < 0$ , and it
is then ruled out.

\section{The QGP phase : $T_E < T$}

Several analytic models have been proposed
to describe the properties of this 
phase and, as analytic
models can be obviously analytically continued in the  $\mu^2 \le 0$
half plane, imaginary chemical potential is an excellent
testbed for these approaches. Here, as an example, we
just contrast the data with a free field behavior
$\Delta P / T^4 = 2(\mu/T)^2\;,$
where $\Delta P = P(\mu) - P(0)$, and we have ignored the
fourth order terms. 
In Fig. 4 we plot $ \Delta P K_{L(N_t=4)}/(T^4 2 (\mu/T)^2) $
versus $\mu/T$, where we have corrected for finite lattice effects
 $K_{L(N_t=4)}$ following \cite{Fodor:2002km}, \cite{Allton:2003vx}.

This ``effective prefactor plot'' 
is perhaps more informative than the direct quadratic fits  
of $\Delta P / T^4$ 
to $k (\mu/T)^2$, 
as it allows an assessment by eye of the $\mu$ dependence, if any, of the
prefactor to the quadratic term. 


We see that the results approach the perturbative limit, but corrections at
small chemical potential are clearly visible,
possibly consistent  with the predictions of \cite{Letessier:2003uj}.

\section{Summary/Outlook}

We have studied four flavor QCD  within
the imaginary chemical potential approach in
a large part of the phase diagram.

We have shown that
the chiral and deconfining transition remain correlated at
nonzero chemical potential. We have found that the critical
line is described by a polynomial,
and interpreted this result in terms of simple models. 
We have identified and discussed three different regimes: 
the hadronic phase results are consistent with the hadron resonance gas model.
The Roberge Weiss regime is eminently
nonperturbative, and in this regime we have the possibility to study
the nature of the chiral transition at nonzero chemical potential
without performing any analytic continuation. The
Quark Gluon Plasma phase is an ideal test bed for analytic models.

The method seems mature for quantitative studies in realistic cases,
and a nice possibility is offered by a combination of this approach
with other methods, for instance by using reweighting \cite{Fodor:2002km}
\cite{Crompton:2001ws}  or direct
calculations of derivatives \cite{Gavai:2003mf} 
at nonzero $\mu$ to improve the accuracy of the results at negative $\mu^2$.
Finally, the study of discontinuities as sketched in Sect.3 above
might offer an alternative approach to the study
of the endpoints and tricritical points.

\end{document}